\documentclass[num-refs]{wiley-article} 

\usepackage{siunitx}

\usepackage{amsmath,amssymb,amsfonts}
\usepackage{algorithmic}
\usepackage{graphicx}
\usepackage{textcomp}
\usepackage{xcolor}

\usepackage{booktabs}
\usepackage{multirow}

\usepackage{hyperref}
\usepackage{subcaption}

\usepackage{microtype}

\usepackage{tikz}
\usetikzlibrary{shapes,arrows,chains,calc}
\colorlet{lcfree}{green}
\colorlet{lcnorm}{blue}
\colorlet{lccong}{red}

\papertype{Draft}
\paperfield{}

 \title{Industrial Control via Application Containers:
        Maintaining determinism in IAAS
        }
        
\author[1,3\authfn{1}]{Florian Hofer}
\author[2]{Martin Sehr}
\author[3]{Alberto Sangiovanni-Vincentelli}
\author[1]{Barbara Russo}

\affil[1]{Faculty of Computer Science, Free University of Bolzano-Bozen, Bolzano, 39100, Italy}
\affil[2]{Corporate Technology, Siemens Corporation, Berkeley, CA, 94704, USA}
\affil[3]{EECS, University of California, Berkeley, CA, 94720, USA}

\corraddress{Florian Hofer, Faculty of Computer Science, Free University of Bolzano-Bozen, Bolzano, 39100, Italy}
\corremail{florian.hofer@stud-inf.unibz.it}

\fundinginfo{iCyPhy, through Siemens Corporation, Corporate Technology}

\runningauthor{F. Hofer \textit{et al.}}

\begin{document}

\maketitle

\begin{abstract}
    \small 
    Industry 4.0 is changing fundamentally data collection, its storage and analysis in industrial processes, enabling novel application such as flexible manufacturing of highly customized products. 
    Real-time control of these processes, however, has not yet realized its full potential in using the collected data to drive further development.
	%
    Indeed, typical industrial control systems are tailored to the plant they need to control, making reuse and adaptation a challenge.
	%
    In the past, the need to solve plant specific problems overshadowed the benefits of physically isolating a control system from its plant.  
    %
    We believe that modern virtualization techniques, specifically \textit{application containers}, present a unique opportunity to decouple control from plants.  
    This separation permits us to fully realize the potential for highly distributed, and transferable industrial processes even with real-time constraints arising from time-critical sub-processes.
    %
    In this paper, we explore the challenges and opportunities of shifting industrial control software from dedicated hardware to bare-metal servers or (edge) cloud computing platforms using \textit{off-the-shelf} technology.
	%
    We present a migration architecture and show, using a specifically developed orchestration tool, that containerized applications can run on shared resources without compromising scheduled execution within given time constraints.
    Through latency and computational performance experiments we explore limits of three system setups and
   	summarize lessons learned.
   	
\end{abstract}

\section{Introduction}

Emerging technologies such as the Internet of Things and Cloud Computing are radically re-shaping structure and control of industrial processes.
These innovations allow the creation of highly flexible production systems, an essential component of the fourth industrial revolution.
Key enabling technologies such as distributed sensing, big-data analysis and cloud storage are taking the central stage in developing new industrial control systems.
Consequently, operation of Edge Computing along with cross-platform features, third party software and mixed criticality applications increase in significance.
The computation requirements given the migration of functionality towards the ``edge'' imply new system architectures~\cite{Telschigetal2018, Hofer2018}. 

The control of industrial processes, however, has not changed much over the last few decades, and there are reasons for it.
Their nature exceeds classical control software requirements by adding further constraints.
Additional requirements towards aspects such as timeliness and ruggedness make maintaining formal guarantees or software changes difficult.
For instance, software for typical industrial processes has to respond to changes in the physical world within predefined time limits.
Moving such control tasks from devices physically co-located with the supervised process to cloud, edge or fog computing platforms, requires dealing with network delays that are difficult to predict.
Moreover, while dedicated hardware and bare-metal solutions may give the control design full authority over the run-time environment,
resource virtualization constrains the determination of a proper environment on cloud computing platforms.
A previously monolithic application is now part of an operating system (OS) managed environment, adding further difficulties such as inter-process or inter-service communication (IPC/ISC).
Yet, we believe that the principles of Industry 4.0 present a unique opportunity to explore complementing traditional automation components with a novel control architecture~\cite{Tascietal2018}.

We believe that modern virtualization techniques such as application containerization~\cite{Mogaetal2016,Tascietal2018,GoldschmidtHauck-Stattelmann2016} are essential for appropriate utilization of cloud computing resources in industrial control systems.
Such techniques would yield the same advantages that traditional containerized micro-services present: the creation of light and easily distributed control applications able to run on any system and that are, at the same time, easy to maintain and update~\cite{Fazioetal2016}.

With control containerization we create a strong enabler for Industry 4.0 attributes.
Beyond the migration capabilities and flexibility, containers simplify the parallel execution of control software on devices such as PLCs and, to a lesser extent, on sensing and actuating field devices.
This results in increased reliability and robustness, while enabling further exploitation of self-* properties (i.e., 
Self-aware, Self-predict, Self-compare, Self-configure, Self-maintain, Self-organize~\cite{Leeetal2015}).
Time-machines (snapshots of control software and/or machine state),
control redundancy (parallel operation of containers and/or virtual server instances~\cite{Leeetal2015})
and online system reconfiguration (reprogramming of control algorithms and product specifications with little or no downtime~\cite{TelschigKnapp2017}) 
are only a few of the Industry 4.0 tools made accessible.
Containers allow applications such as performance and distributed health monitoring~\cite{Wuetal2017,Terrissaetal2016} to run on a shared end node.
They can host a Digital-Twin~\cite{Schroederetal2016} to predict malfunction, maintenance intervals and tool lifespan.
Lastly, these modern virtualization techniques enables mixed criticality contexts, promoting increased efficiency, reduction of the operational cost and decrease of production downtime~\cite{Royetal2016}. 
In this paper, we explore the feasibility of relocating real-time control applications, using off-the-shelf technology, from dedicated infrastructure and hardware onto a shared resource environment, both on a bare-metal host and in the cloud.
The contributions of this paper are: 
\begin{itemize}
    \item An architecture proposal to ease migration and enable extension with, and integration of, Industry 4.0 features.

	\item A proof of concept of an orchestration solution, opt to statically allocate and monitor containers and their resources.
	
	\item Evaluation and resource efficiency tests of the hard real-time task scheduling with application containers
	
	\item Demonstration of how, under specific conditions, the same tasks can be run in the cloud.
\end{itemize}

The rest of this paper is structured as follows. Section~\ref{sec:litmot} analyzes related work and background motivating our analysis.
Section~\ref{sec:arch} proposes an architectural solution, while in Section~\ref{sec:method}, we discuss the methodology and design of experiments.
We next detail the determined run-time contexts for containers, their frameworks, candidate host OSs and system latency in Section~\ref{sec:virtsust}. 
Section~\ref{sec:orchestr} describes orchestration of containers, including software tool and tests.
Finally, we discuss lessons learned and conclude in the last two Sections.

\section{Literature and Motivation}
\label{sec:litmot}

The proposed migration deals with two different areas: high performance computing (HPC) and control software containers.
Both focus on different aspects of control program execution.
The former focuses on lowering its latency and gives less importance to its execution determinism.
The latter tries to reshape its run-time environment and thus, to create a level of independence to its underlying hardware. 
During this redesign determinism stays in focus, leaving system virtualization in the background.
The resulting combination of \textit{containers executed on cloud resources} and \textit{strictly time-dependent control application containerization} constitutes a new challenge that can be coped with applying insights from both fields.
Such a combination requires an operating system kernel that supports and \textit{exceeds} soft real-time guarantees secured by low latency kernel flavors in use on HPC installations while keeping only limited environmental control.
In this paper we assess the feasibility of this approach using off-the-shelf technology.

The following subsections detail motivation and related work.
We discuss briefly useful insights for our investigation and outline the motivation to examine our problem.
The section closes with the research questions for this study.

\subsection{Control containerization}

Containerizing control applications has been discussed in recent literature. 
Moga \textit{et al.}~\cite{Mogaetal2016}, for instance, presented the concept of containerization of full control applications to decouple the hardware and software life-cycles of an industrial automation system.
Due to the performance overhead in hardware virtualization, the authors state that OS-level virtualization is a suitable technique to cope with automation system timing demands.
They propose two approaches to migrate a control application into containers on top of a patched real-time Linux-based operating system: 
a) a given system is decomposed into subsystems, where a set of sub-units performs a localized computation, which then is actuated through a global decision maker, or 
b) Devices are defined as a set of processes, where each process is an isolated standalone solution with a shared communication stack, and based on this, systems are divided into specialized modules, allowing a granular development and update strategy. 
The authors demonstrate the feasibility of real-time applications with containerization, even though they express concern on the maturity of the technical solution presented.

Goldschmidt and Hauk-Stattelmann in~\cite{GoldschmidtHauck-Stattelmann2016} perform benchmark tests on modularized industrial
Programmable Logic Controller (PLC) applications.
This analysis examines the impact of container-based virtualization on real-time constraints.
As there is no solution for legacy code migration of PLCs, the migration to application containers could extend a system's lifetime beyond the physical device's limits.
Even though tests showed a worst-case latency in the order of $15ms$ on Intel-based hosts, 
the authors argue that the container engines may be stripped down and optimized for real-time execution.
In a follow-up work, Goldschmidt \textit{et al.}~\cite{Goldschmidtetal2018}, a possible multi-purpose architecture was described and tested in a real-world use case.
The results show the worst case latency of about $1ms$ for a Raspberry PI single-board computer, making the solution viable for cycle times of about $100ms$ to $1s$.
The authors state that topics such as memory overhead, containers' restricted access and problems due to technology immaturity are still to be investigated.

Tasci \textit{et al.}~\cite{Tascietal2018} address architectural details not discussed in~\cite{GoldschmidtHauck-Stattelmann2016} and \cite{Goldschmidtetal2018}.
These additions include the definite run-time environment and how deterministic communication of containers and field devices may be achieved in a novel container-based architecture. 
They proposed a Linux-based solution as host operating system, including both single kernel preemption-focused PREEMPT-RT patch and co-kernel oriented Xenomai. 
With this patch, the approach exhibits better predictability, although it suffers from security concerns introduced by exposed system files required by Xenomai. For this reason, they suggested limiting its application for safety-critical code execution. 
They analyzed and discussed inter-process messaging in detail, focusing on the specific properties needed in real-time applications.
Finally, they implemented an orchestration run-time managing intra-container communication and showed that task times as low as $500\mu s$ are possible.


The three solutions discussed above share one common aspect: \emph{they base on a bare-metal configuration}. 
These solutions illustrate a first step for the re-allocation of an embedded control software onto a dedicated infrastructure.
They all consider real-time constraints but remain limited to the execution on physical hardware.
However, a take-way remains that containerization of hard real-time applications is viable. 

\subsection{Cloud and High Performance Computing}

In 2014, Garcia-Vallas \textit{et al.}~\cite{Garcia-Vallsetal2014} analyzed challenges for predictable and deterministic cloud computing.
Even though they focus on soft real-time applications, certain aspects and limits apply to any real-time systems.
Merging cloud computing with real-time requirements is a challenging task; the authors state the guest OS has only limited access to physical hardware and thus suffers from unpredictability of non-hierarchical scheduling, and thick stack communications. 
While there exist real-time enabled hypervisors that manage virtual instances such as the paravirtualized RT-Xen with direct access to hardware, the shared resources still suffer from latency that may make real-time execution impossible.

Hallmans \textit{et al.}~\cite{Hallmansetal2015} draw similar observations, but they reach different conclusions.
They not only conclude that it is possible to move a complete soft real-time system into the cloud, the authors see an upcoming
development that further allows for hard real-time systems.
Many latency performance evaluations confirm this possibility.
Nonetheless, to our knowledge no one has verified the proper execution of real-time tasks within deadlines.

\subsection{Architecture and Scheduling}

Felter \textit{et al.} in~\cite{Felteretal2015} focused on identifying the performance of instances based on hardware virtualization via Kernel-based Virtual Machines (KVMs) and container OS-virtualization using the cross-platform capable Docker. 
The benchmarks confirm that Docker results in equal or better performance than KVMs in almost all cases.
Arango \textit{et al.}~\cite{Arangoetal2017} analyzed three containerization techniques for use in cloud computing. 
The paper compares Canonical's Linux Containers (LXC), Docker and Singularity, an engine developed by Lawrence Berkeley National Laboratory, to a bare-metal application. 
In many aspects, the Singularity containers performed better, sometimes even better than the bare-metal implementation, but this is largely due to the blended approach of the engine; Singularity is an incomplete virtualization solution since it grants access to I/O operations without context changes.

A recent work by Telschig \textit{et al.}~\cite{Telschigetal2018} explores a platform-independent container architecture for real-time systems.
The authors identify mixed-criticality, cross-platform operation and third party software use as main reason for the development of new architectures.
In their proposal manages communication between this dependent distributed software through an architecture.
This architecture focuses on isolation of critical from non-critical tasks and portability.
The presentation concludes with the introduction of a prototype agent.

Abeni \textit{et al.}~\cite{Abenietal2018} tried to extend the Linux standard scheduler to get better response times.
In their recent work they detail how to extend the complete fair scheduler (CFS) hierarchically with a deadline based algorithm optimizing latency results for containerized software.
The modified scheduler successfully manages larger amount of time critical tasks, performing better than the default deadline based scheduler.

Ultimately, containerization has shown powerful enough to be a resource economic replacement for traditional virtualization techniques.
However, a performance investigation with real-time applications remains due; scheduling techniques like those presented in~\cite{Abenietal2018} further prove that there is still room for improvement.

\subsection{Research question and motivation}
\label{sec:bgnd}

While flexibility and efficiency are big advantages of the new paradigm, such Smart systems display increased running costs. 
New architectures suggested for the fourth industrial revolution mostly provide for distributed and, above all, decentralized control.
This control layer however has to carry out increasingly complex tasks.
With the resulting high amount of small distributed supervision loops grows the need of dedicated hardware performing a single task.
In turn, this would increase maintenance and operation costs.

Virtualization of control units, i.e., abstracting the function from hardware, allows up-scaling the installed computation appliances.
Such unit can run on shared hardware exploiting cost reduction advantages typical to cloud computing environments.
To keep a low maintenance profile, this up-scaling has to use standard hardware and software.
The goal of our experiments is therefore to explore whether, and to which extent, off-the-shelf technology can help to migrate hard real-time applications to a virtualized computing resource.
After successful migration, practitioners can run applications on a smaller, centralized amount of computing entities, consequently saving resources and substantially reducing operational cost.

A requirement for success remains that the software keeps its timing within bounds past migration.
Control software usually characterizes by one or more real-time tasks with periodic execution and a computation deadline.
In the literature, three categories of periodic real-time applications have been analyzed:
\emph{Soft}, where computation value decreases with a deadline overshoot;
\emph{Firm}, when exceeding the maximum delivery time nulls the computation value; 
or \emph{Hard}, where a missed deadline may have catastrophic consequences~\cite{Buttazzo2011}.
A task that exceeds its timing limits may further impede the execution of dependent and independent tasks. 
The delayed scheduler yield takes additional resources that may cause a bottleneck and following tasks may not maintain their deadlines.
Consequently, if running multiple real-time capable instances, we have to verify that all task follow their run-time parameters.

A single relationship identifies all these parameters.
The total required computation time $c_i$ of a periodic real-time application $i$ relates to its relative deadline $d_i$ and period $p_i$ in the following manner:
\begin{equation}\label{eq:deadline}
	c_i = f_i + r_i = \textcolor{orange}{f_i} + \textcolor{brown}{\sum_{j=1}^{N} n_{ij}} + \sum_{k=1}^{M} n_{ik} + t_i \leq d_i \leq p_i
\end{equation}
where $f_i$ is the wake-up or firing time (latency) and $r_i$ is the total run-time. 
The former captures the time spent between the period start and the execution start of a task.
Its measurement includes task switching times and delays due to higher priority task and interrupts served.
The latter expresses the actual used computation time $t_i$ and task ($n_{ik}$) or environment ($n_{ij}$) induced noise.
This task noise includes interruptions by higher priority tasks, task IPC and I/O waits and latency due to missed pages, while environment noise includes hypervisor delays and hardware or (kernel) software interrupts.
If the sum off all factors ($c_i$) exceeds the relative deadline $d_i$, the resulting misbehavior of a controlled system might have catastrophic consequences.
Hence, monitoring and containing these parameters can make a migration sustainable.


\begin{figure}[tb]
		\centering
		\includegraphics[width=0.8\linewidth]{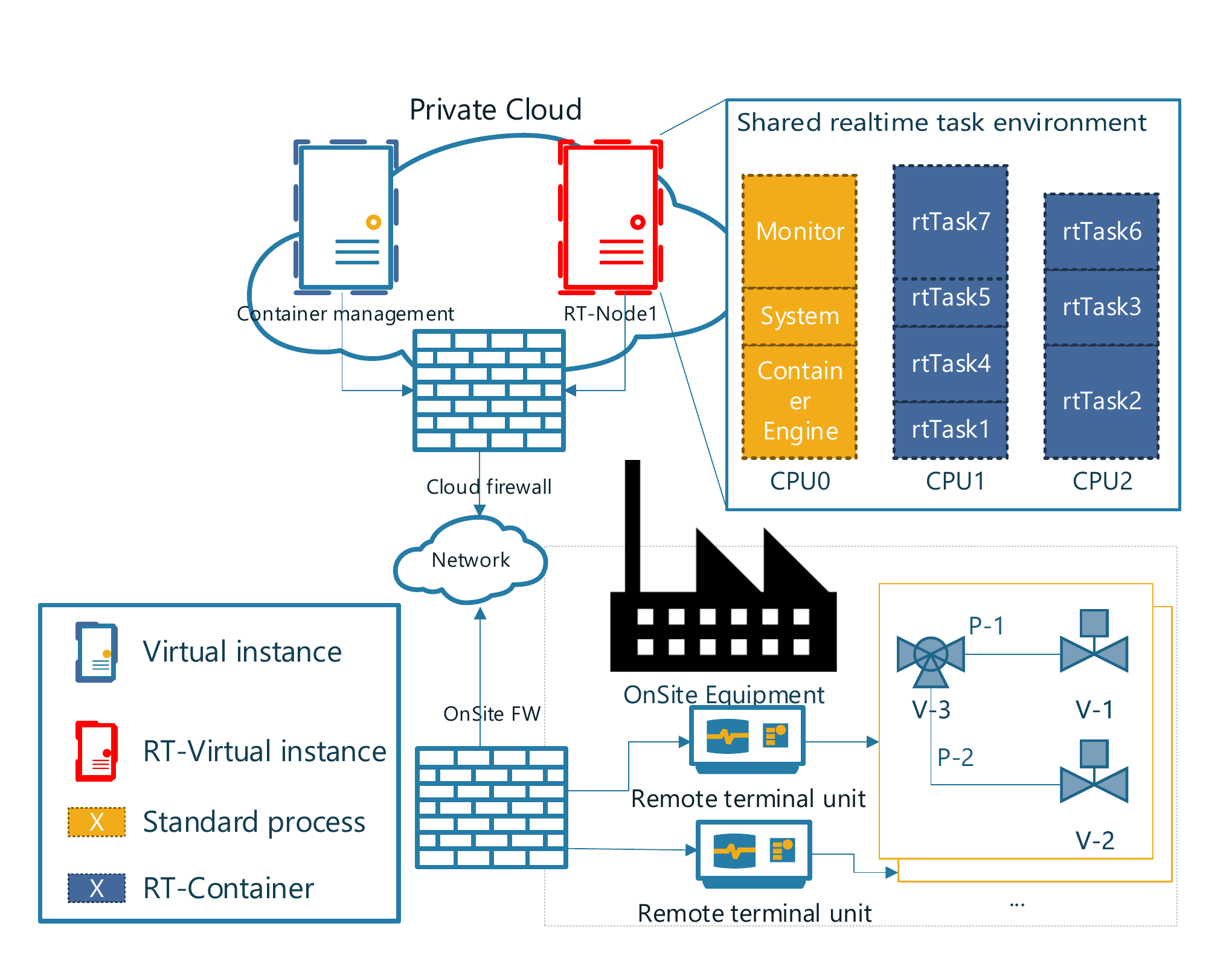}
		\caption{Motivating example: A cooling and auxiliary regulation system configuration for a gas turbine migrated with to real-time enabled cloud. RT-Node1 monitors and handles the on-premises installation.}
		\label{fig:smpcase}
\end{figure}


\textbf{Motivating Example: }
Figure~\ref{fig:smpcase} depicts a distributed facility that controls the auxiliary and cooling systems of a group of thermo-electric gas-turbines.
It illustrates an example of a migration to a software system with shared resource and application containerization.
Like in Hallmans ~\textit{et al.}~\cite{Hallmansetal2015}, the control components are separate from the on-site remote terminal unit and run in a common, two instance virtualized environment.
The real-time capable virtualization instance, right in the private cloud, acts as an intermediary between ``Monitoring and Management'' and the on-premises end-terminals.
Due to migration, this architecture abstracts the control logic from the production site requiring a reorganization of its software.

When migrating to a shared-resource system we want to assign each application binary to its dedicated environment.
As suggested by Moga \textit{et al.}~\cite{Mogaetal2016}, the software has been divided into multiple independent binaries, isolated and adapted to run on a standard Linux system.
The application refresh rate, or periodicity $p$, should not exceed the expected maximum round-trip time between remote terminal unit and cloud of $100ms$.
If we assume now the system software exhibits the worst case computing time (WCET) of $10ms$ and that each instance uses assigned CPU and memory exclusively, 
the remaining CPU time of $90ms$ is spent in idle mode, resulting in high resource waste.
Sharing the spare CPU-time can reach a better resource utilization. 
Placing multiple containers on the same resources can additionally reduce the required system size and its running costs.
Separate running environments enable thus flexible resource management and may reduce infrastructural cost.


\noindent To verify migration behavior, we examine the following research questions:
\begin{enumerate}[label=\textit{RQ\arabic*:},start=1]
	\item \textit{What are possible off-the-shelf system configurations that make resource sharing through containers viable?}\\
	We have seen in Equation~\ref{eq:deadline} that the achievable amount of resource sharing depends on the system and concurrent running tasks.
	This research questions intends to investigate the responsiveness of possible candidate off-the-shelf systems.
	Systems that prove a low and stable firing time, $f_i$, fulfill a vital prerequisite to achieve determinism in a shared context. 

	\item \textit{What is the achievable level of CPU sharing with a standard real-time enabled kernel?}\\
	While a constant $f_i$ describes vigor of reaction, actual CPU load will show the variability of task run-times, $r_i$.
	Monitoring software run-time on shared CPUs thus displays the impact of task interaction and operating system, I/O and virtualization delays on the programmed run-time $t_i$, and finally, the determinism.
	By isolating system ($n_ij$) from task noise ($n_ik$), we can determine the upper bound of a system's availability to resource sharing.
	
\end{enumerate}

In this paper, we approach these two points and focus on the feasibility of a migration.
Firstly, we illustrate latency behavior with shared resource on multiple hardware and system setups. 
Exploring operating systems and container engines, we select configurations for latency stress tests. 
A monitored test will show how task reaction times change with varying configurations and system load, RQ1.
Secondly, we extend the experiments on the best performing candidates to analyze performance and determinism with different loads. 
Through static resource allocation we can further explore computation time stability, delays and occurring deadline misses if more than one task runs to the same resource.
Isolated tests of CPU performance allow us to remove confounding factors and have a hint on the upper sharing boundary, RQ2.
Section~\ref{sec:tests} will tackle the answers of these research questions.

The motivating example of this section further shows that the migration of control requires adaptation and reorganization of software and system structure.
To ease such migration, we extend our investigation with an architecture proposal for the Industry 4.0 context.
Its design acts as template easing transition to a containerized setup and shared instances, and it enables advanced features for novel industrial control systems.
The next section illustrates and details this architecture proposal.

\section{Real-Time Smart System Architecture}
\label{sec:arch}

Migrating from dedicated hardware to shared resources requires more than a simple relocation strategy.
The control algorithm needs adaptation and reorganization into units that 
take care of the different responsibilities of a control system within their constraints and run-time parameters~\cite{Hallmansetal2015}.
A ``leveled'' design~\cite{Hallmansetal2015} can split the responsibilities of a such an algorithm depending on their criticality and timeliness, enabling the integration of cloud systems into a real-time industrial process.
Via a layered approach to system architecture we perform this allocation of responsibilities for a smart system.

The challenges of a migration do not end on a system level.
Once a practitioner adapts and configures a binary to run on the new system, he or she has to monitor the correct execution within its timing parameters.
The interaction among control applications and unmanaged inter-process communication may cause irregular and unpredictable run-time behavior~\cite{Tascietal2018}.
Also, the amount of real-time applications potentially running concurrently on a single node requires for adequate monitoring and management tools to avoid overloads or mutual influence. 
Thus, the system architecture proposed in this section supports the migration of control applications on shared resources (i.e., \textit{control virtualization}), aiming at addressing the above issues.

\subsection{Overview and Layering}


\begin{figure}[bt]
	\centering
	\includegraphics[width=\textwidth]{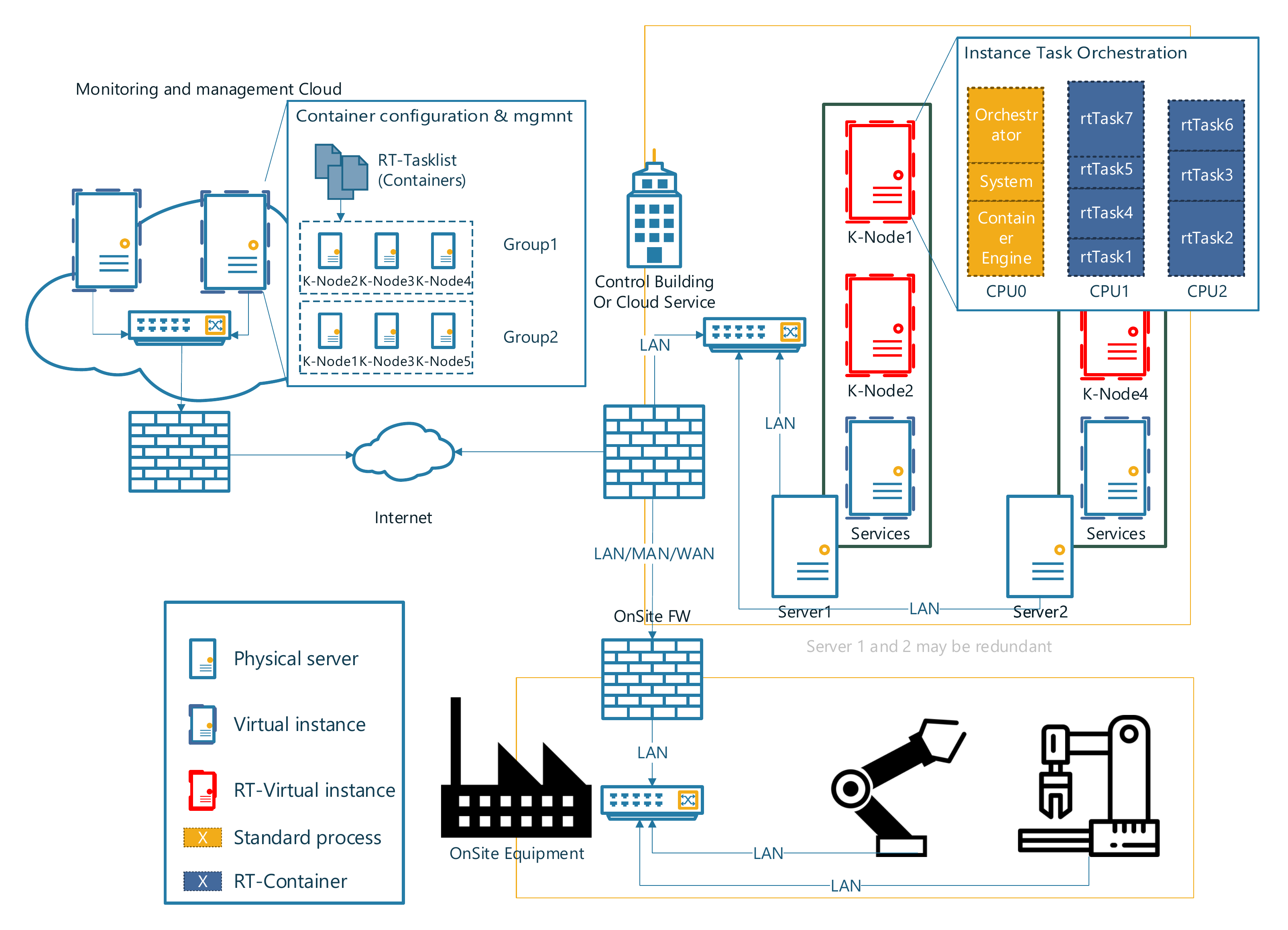}
	\caption{
		Proposed architecture for Container-based Virtualized Industrial Control Systems. 
		The Monitoring and Management Cloud monitors the system and deploys containers in the Control Cluster.
		The physical servers dedicated for control tasks operate multiple virtual machines, each hosting several real-time application containers. 
		It is the responsibility of the Orchestrator to organize and monitor the run-time load.}
	\label{fig:sys-arch}
\end{figure}

Our architecture extends the concept of the ``leveled'' model of Hallmans \textit{et al.}~\cite{Hallmansetal2015} to off-the-shelf technology and managed real-time systems fitting the requirements of virtualized control software.
We divide the architecture into three layers (Figure~\ref{fig:sys-arch}):
\begin{itemize}
	\item The \textbf{Monitoring and Management Cloud} or ``Cloud'' in Hallmans \textit{et al.}, i.e., services hosted on cloud or private virtualized dedicated infrastructure;
	\item The \textbf{Control Cluster} or ``Real-time Cloud'', for process control and control-related services;
	\item The \textbf{On-premises Installation} or ``Process'', connecting a multitude of heterogeneous devices that interact with the physical world.
\end{itemize}
The  three layers may overlap such that, for example, the control cluster can be part of the cloud or on-premises installation as an internal IAAS infrastructure.

\begin{figure}[bt]
	\centering
	\includegraphics[width=0.9\textwidth]{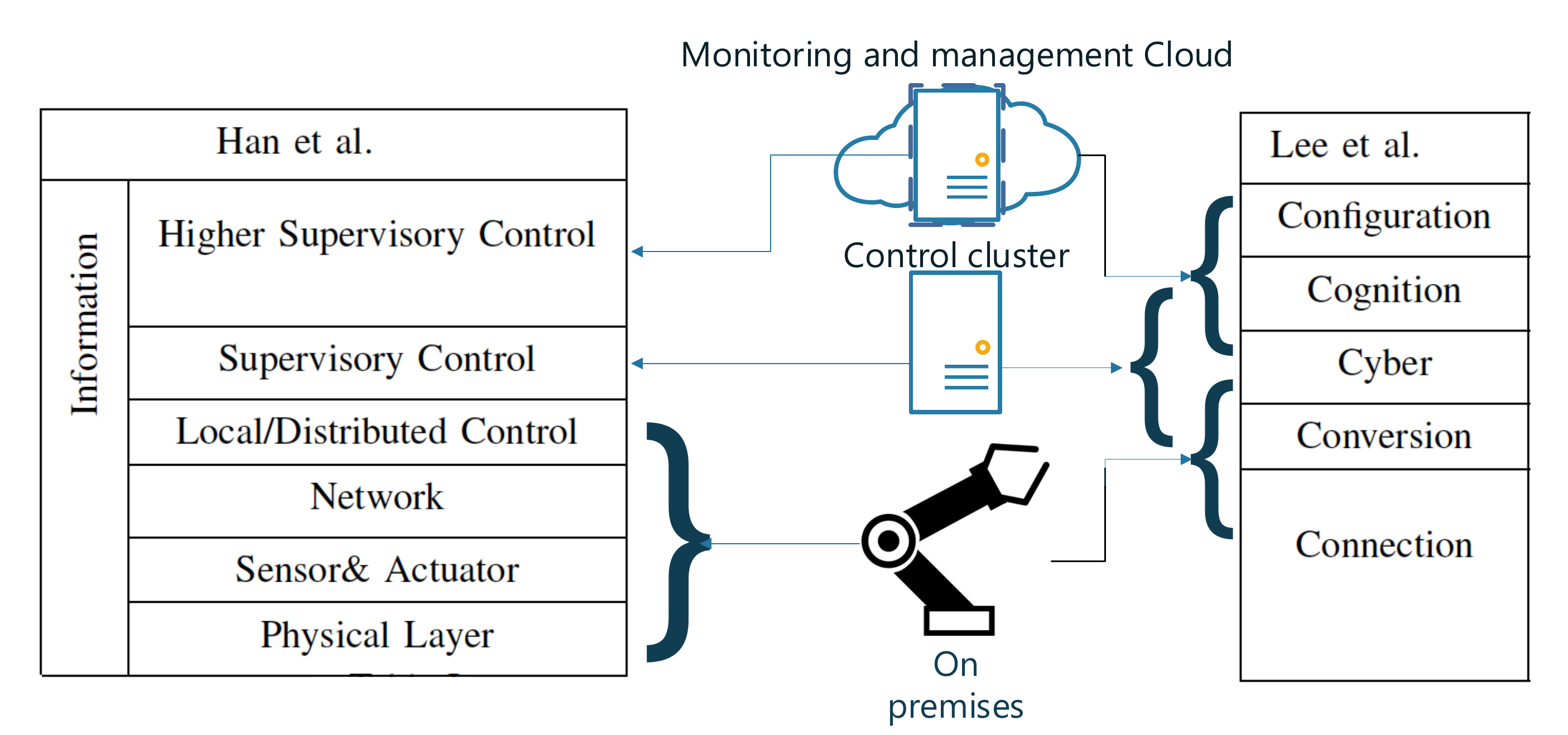}
	\caption{Comparison and mapping of layers to the architecture proposal of the two used references, Han \textit{et al.}~\cite{Hanetal2014} and Lee \textit{et al.}~\cite{Leeetal2015}.}
	\label{fig:layers}
\end{figure}

We use then the layer classification and analysis technique of recent work~\cite{HoferRusso2020} to integrate the model with two further styles.
The first integration is the 5C model of Lee \textit{et al.}~\cite{Leeetal2015}.
This model represents a hierarchical distribution of functions, divided into five layers, each representing Industry 4.0 attributes.
This enables us to locate attributes such as \emph{self-compare} or \emph{self-optimization} in our proposal.
The three control levels described by Han \textit{et al.} act as an incentive for control loop division and ease vulnerability investigations~\cite{Hanetal2014}.
These loops, i.e. Local Control, Supervisory Control and Higher Supervisory Control, help the placement of algorithm components. 
This reflects the division by criticality and timing requirements mentioned by Hallmans \textit{et al.}, setting up three loop control levels. 
Figure~\ref{fig:layers} illustrates mapping of layers and competences.

As we illustrate in the rest of this paper, this architecture allows for a better implementation of virtualized control. 
Through a layered approach we ease problem identification and handling. 
It enables detailed assessments such as gradual detection of security issues and adaptation of architectures as carried out in within similar heterogeneous environments~\cite{Hofer2018-2,HoferRusso2020}.
In the following sections, we detail the three layers and their mapping to the two reference architecture styles while pointing out the connection to functions and attributes of Industry 4.0, refereed as~\emph{I4.0}.

\subsection{Management and Monitoring Cloud}
\label{sec:cloud}

The first component of the architecture consists in cloud-based monitoring and management infrastructure and services, top left in Figure~\ref{fig:sys-arch}.
Many of the architectural approaches introduced after publication of the Industry 4.0 vision include this component as a hub.
In this layer, data is globally collected and analyzed and data-dependent supervision decision are taken.
It performs data acquisition and aggregation from the on-premises devices and analyzes them, for instance, through artificial intelligence tools.
The integration in the layer of distributed diagnosis and prognosis frameworks as proposed in Wu \textit{et al.}~\cite{Wuetal2017} allows for host machine learning processes based on collected and aggregated plant data.
Techniques such as Preventive Health Management (PHM) as a Service, which reduces the maintenance effort for the plant operator by relying on Platform as a Service (PaaS) and Software as a Service (SaaS) can be implemented ~\cite{Terrissaetal2016}.
Such frameworks and services ultimately enable self-adjustment and self-optimization techniques to reduce production waste and adapt to variations such as mechanical wear (I4.0 ``Configuration'' level). 

This cloud layer also hosts a service for container management, providing instruments for appropriate planning, positioning, and execution of real-time containers.
Real-time tasks as well as containers are arranged according to their function and interdependence and deployed on available real-time capable nodes, called K-nodes in Figure~\ref{fig:sys-arch}.
Through the help of client side agents, the service is able to seamlessly update and replace distributed applications during runtime.
This component enables self-configuration (I4.0 ``Configuration'' level) by taking care of the software replacements based on a reconfiguration plan~\cite{TelschigKnapp2017}.
Paired with the container management tool, a system monitoring tool can verify the container execution.
Differently from the above-cited PHM and diagnosis and prognosis frameworks, the aim of this monitoring tool is not production surveillance, but monitoring the health status of K-nodes and virtual servers.
Extension of the monitoring service with a time series database further allows tracking changes in time, performing data analysis, and applying data-based techniques such as deep learning.

A further service placed at this level of architecture is an interface to the human operator as
plant operators in an Industry 4.0 context interact with the system in a more decision-making than an operative role. 
As such, information is displayed to enable informed operators to make decisions and interventions in production processes.
To this aim, features such as simulation and synthesis may optionally be available~\cite{Goreckyetal2014} (I4.0 ``Cognition'' level).
Finally, replicas in form of a ``digital twin'' test reliability and the overall monitoring and supervision of the cloud environment (I4.0 ``Cyber'' level).

\subsection{Control Cluster}
\label{sec:control}

The central element of the architecture is an IAAS infrastructure with main purpose to host services and processes that have to interact with on-premises devices.
Figure~\ref{fig:sys-arch} shows an example for a hardware configuration for container-based virtualized industrial control systems.
Depending on the system's needs, the represented components are either virtual servers (or instances) running in a cloud or physical servers operated in a private (edge) cloud. 
In the latter case, each server can run more than one virtual instance, obtaining again the same resource sharing advantages of a computing cloud infrastructure.
In both cases, the hosted virtual instances can be real-time capable running control software or non-real-time capable for further services.
The real-time instances, called K-nodes in the Figure, are running multiple containers managed by the cloud service.
A dedicated tool orchestrates system resources at run-time (See Section~\ref{sec:orch}).
In this environment, each binary of an application can be managed within one container to which we can add constraints and boundaries to ensure operations.

As noted by Telschig \textit{et al.}~\cite{Telschigetal2018}, the continuous growing demand on extension of control loops with cloud based analytics (see Section~\ref{sec:cloud}) requires mixed-critical software components to run on the same system.
Thus, in the Control Cluster, a time critical component runs on fix assigned resources to guarantee timeliness.
It is isolated from the components that run on a best-effort CPU scheduling policy, while still sharing resources of the system.
In this setting, non-real-time instances or separately allocated resources can handle such best-effort tasks.
The co-located best-effort resources may then be reclaimed to buffer for real-time task resource shortage.
Non-real-time instances can then handle other, less critical, services.
For instance, they can run a time-machine or carry the edge computing portion of the health monitoring framework detailed in Wu \textit{et al.}~\cite{Wuetal2017} (I4.0 ``Conversion'' level).
The former collects snapshots of real-time applications to enable peer comparison and similarity analysis, thus promoting self-awareness~\cite{Leeetal2015}.
The latter operates with redundancy on multiple copies of containers (I4.0 ``Cyber'' level), or the virtual instances themselves can have replicas to increase system's robustness.
Server 2 in Figure~\ref{fig:sys-arch} could be a replica of Server 1, ready to take control when the latter fails.
We can have replicas in form of ``digital twins'', 
requiring the real-time application to be extended by a model representing the device and its environment.
The model, fed with sensory input coming from on-premises and interfaced with the running process and/or human operators, finally allows further self-comparison and diagnosis~\cite{Schroederetal2016}.

\subsection{On-premises Installation} 

The control software connects with the sensing and actuation devices placed on or near the equipment of the factory (I4.0 ``Connection'' level).
Depending on the timing and determinism requirements, this connection might need to follow more restrictive protocols.
An example of such protocols can be found in the Time Sensitive Networking (TSN) standards family~\cite{ieee01}.
However, application-specific needs and physical location set the need of such protocols.
Depending on control requirements including device distance and cycle times, popular COTS Ethernet enabled protocols may suffice.
Traditional choices such as isochronous ProfiNet and EtherCAT manage hundreds of devices in time-critical manner for local networks~\cite{Robertetal2012}.
The proposal leaves thus the choice of the connection type to each application case.

Although the on-premises computation has been moved to the Control Cluster component, the proposed style foresees further control software installed on on-premises devices as well.
For redundancy purposes, the cluster may indeed operate a redundant copy of the on-premises controller (Section~\ref{sec:control}), or some units may operate as Remote Terminal Units (RTU), serving as interface to the containerized software or even execute some minor local control function.
Such local control loops~\cite{Hanetal2014} would have the advantage to reduce latency while exploiting the computing power of a (private) cloud.

Morabito~\cite{Morabito2017} shows that control applications can run inside a container on typical ARM single board computers with minimal performance impact. 
Replication and snapshots further enable Industry 4.0 features also for such on-site devices.
As part of data evaluation and sharing, they can now independently calculate health, estimated remaining useful life etc., bringing \emph{self-awareness} to machines.
Via snapshots, a machine can compare its performance with itself and others of a fleet enabling \emph{self-comparison}~\cite{Leeetal2015}.
Thus, through containerization we ease maintenance and reduce cost while increasing resilience and robustness.

\subsection{The Orchestrator}
\label{sec:orch}


The heart of this proposed architecture style is the orchestration software running on each real-time capable node of the Control Cluster.
An \emph{orchestrator}, in this context, is a tool developed to increase resource utilization without significantly impacting determinism.
It monitors containers and resources, and assigns the latter according to algorithms, rules or predetermined configurations.

There are two ways to manage resources: \textit{static} and \textit{dynamic}.
%
If statically configured, the level of latency and determinism that is achievable can be defined up-front. 
A static resource schedule is created off-line and passed to the orchestrator for execution.
Although such a configuration would be the safest, the amount of resource sharing gained is limited.
For such a static schedule, the configuration must be pessimistic, taking worst case execution times as regular and granted, and reserving the corresponding CPU-slice for every application.
For higher resource savings, a dynamic reallocation strategy is attractive.
A dynamic scheduling strategy instead reallocates containers during run-time to guarantee timeliness when unforeseen delays occur.
It allows higher resource sharing as it can adapt to current needs.
However, complete dynamic rescheduling of containers would be non-deterministic as it depends on the feasibility/admission test~\cite{Buttazzo2011}.
With given constraints, the determinism can however be managed within a certain probability of success.

In dynamic resource management, instead of allocating resources based on worst case parameters, it uses probabilities to asses situation~\cite{Hoferetal2020}. 
The orchestrator considers typical run times, contemporaneity factors and probabilities of occurrence of the WCET.
It samples run-times and performs curve-fitting to predict distribution models and probabilities.
The combined probabilities then tell the rate of success of a schedule and trigger resource organization as needed.
This approach resembles the ``vertical scaling techniques'' used in cloud-hosted applications~\cite{Shekharetal2018}.
Similar approaches in cloud computing environments increase resource efficiency through over-subscription where the reserved resources may exceed the actual requirements acting as buffers for worst case situations~\cite{Chenetal2018}.
In our case, can be assessed to which probability a system-wide malfunction may occur, allowing a system administrator to set a maximum acceptable boundary of risk. 
This boundary then defines the probability of success of dynamic scheduling in relation with the achieved resource savings: the higher the risk, the more savings may be achieved.

\subsection{Summary}

Migrating control applications from hardware to bare-metal and finally to an IAAS infrastructure has two major advantages.
First, application containerization allows managing and monitoring execution easily. 
It eases parallel operation, redundancy, quick updates and upgrades for container-confined code.
We have seen that replacing a set of running containers at run-time is feasible and it allows distributed updates and life-fixes of critical problems. 
A container may also execute on the on-site device while keeping a copy on the IAAS for backup and/or redundancy.
Second, physical hosts, in cloud and private cloud, serve multiple virtual instances. 
The computing power available to individual instances is often flexible.
Such computing power usually exceeds the original hardware's performance, permitting us to use more complex and demanding control algorithms.

On the other hand, the process of application containerization requires some software adaptation, generalization of interfaces and I/O, and results consequently in hardware abstraction.
This extra effort must be taken into account when evaluating a migration strategy.
In addition, the introduced distance between control cloud and on-site devices may require using more time critical communication protocols respecting standards such as the TSN family, which may add some overhead. 

On the other side, a recent systematic mapping study~\cite{Hofer2018} highlights the limits of available architectures, for instance based on the 5C attributes. 
The architecture style illustrated in this section has been designed  not only to ease application migration, but also to support self-* properties of the migrated application with management and monitoring layer.
In summary, the proposed architecture gives support to a more complete control solution for both research and industry.

\section{Methodology and Design of Experiments}
\label{sec:method}
\begin{figure}
	\centering
	\begin{tikzpicture}[%
		>=triangle 60,              
		start chain=going below,    
		node distance=6mm and 40mm, 
		every join/.style={norm},   
		scale=0.7,
		every node/.style={transform shape}
		]
		\tikzset{
			base/.style={draw, on chain, on grid, align=center, minimum height=4ex},
			proc/.style={base, rectangle, text width=8em},
			test/.style={base, diamond, aspect=2, text width=5em},
			term/.style={proc, rounded corners},
			nmark/.style={draw, cyan, circle, font={\sffamily\bfseries}},
			norm/.style={->, draw, lcnorm},
			free/.style={->, draw, lcfree},
			cong/.style={->, draw, lccong},
			it/.style={font={\small\itshape}}
		}
		\node [term] (s0) {Experiment start};
		\node [proc, right=of s0, join] (p0) {Evaluation of Candidate OS};
		\node [proc, join] (p1) {Evaluation of Containerization techniques};
		
		\node [proc, right= of p0] (p2)	{Virtualization settings test};
		\node [proc, join] (p3)	{Test execution};
		\node [proc, join] (p4) {Analysis of resutls};
		
		\node [proc, right=of p2] (p9) {Kernel settings test};
		\node [proc, join] (p6) {Test execution};
		\node [proc, join] (p8) {Analysis of resutls};
		
		\node [term, join, right=of p8] (p10) {Experiment end};
		
		\node [above=0mm of p0, it] {Context};
		\node [above=0mm of p2, it] {Latency};
		\node [above=0mm of p9, it] {Performance};
		
		\draw [->,lcnorm] (p1.east) -- ++(7mm,0mm) |- (p2);
		\draw [->,lcnorm] (p4.east) -- ++(7mm,0mm) |- (p9);
		
		\draw[orange,thick,dotted] ($(p0.north west)+(-3mm,5mm)$) rectangle ($(p4.south east)+(3mm,-3mm)$);
		\draw[brown,thick,dotted] ($(p3.north west)+(-3mm,3mm)$)
		-| ($(p9.north west)+(-3mm,+5mm)$)
		-| ($(p6.south east)+(+3mm,-3mm)$)
		-| ($(p3.north west)+(-3mm,+3mm)$);
		
		\node [brown,right=5mm of p9, it] {\shortstack{Same Hardware and\\ VM systems in all tests}};
		\node [orange,below=5mm of p1, it] {\shortstack{Hofer et al. 2019}};	
	\end{tikzpicture}
	
	\caption{The flowchart shows the steps of the experiment protocol for the study of  performance and latency including the approach of Hofer et al.~\cite{Hoferetal2019}}
	\label{fig:flowchart}
\end{figure}
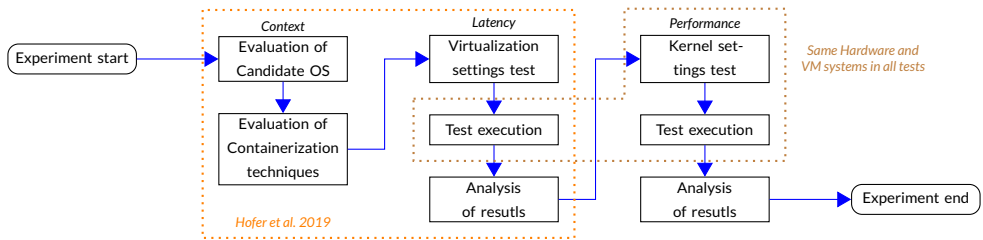

To proceed with a migration onto IAAS, we require experiments that confirm the viability of a migration, Figure~\ref{fig:flowchart}.
These experiments have to validate adequacy of system latency and the sufficiency of run-time determinism.
In Hofer \textit{et al.}~\cite{Hoferetal2019} we explore the running context and execute qualifying latency tests to assess system and Hypervisor influence.
Section~\ref{sec:virtsust} summarizes experiment setup and findings. 
To measure determinism we evaluate in further experiments how a real-time constrained task behaves in a shared and containerized environment.
New performance tests evaluate additional system configurations and assess CPU sharing limits while a specifically developed orchestration tool monitors the environment.
Their results will display the possibility and sharing limits of a IAAS based control infrastructure, on bare-metal and virtualized systems.


Specifically, we compare bare-metal with virtualization approaches that use hypervisors of Type 1 (native).
The former is expected to perform better in latency, but worse in resource economy whereas the latter display better resource economy, but limited in hardware control.
%
Each system will run the same Real-Time enabled OS and test software that will log measurement data during run-time. 
We compare resource sharing capabilities on the following three architectures: 
\begin{itemize}
	\item A bare-metal server, which we use as migration baseline for a typical industrial control systems; 
	\item A Type 1 hypervisor controlled virtual generic instance;
	\item A Type 1 hypervisor controlled virtual compute-optimized instance.
\end{itemize} 

The latency tests verify the suitability of specific hardware or virtualization solutions.
By applying computational and I/O stress to a task's shared resources we can examine latency effects on its real-time parameters.
We start by measuring firing time variations on a system with Type 2 hypervisor to identify the best performing virtualization setup in both cases, when idle and stressed (settings test).
During test run we gradually isolate the measurement tasks, the guest OS and host OS using tools like Linux control groups (\textit{CGroups}) and system configurations such as task and interrupt affinity. 
We pick the best configuration based on low standard deviation (stability) and reduced firing time (reactivity).
Then, we perform the same test with the best configuration on the three mentioned architectures.
We track on all tests how the latency parameters alter as we change the environment and pick again best performing configurations based on stability and reactivity.
The analysis of the data will then allow to make a judgment on the level of suitability of each architecture.

In the second part we focus on the interaction of virtualized control tasks with the shared environment.
The performance tests execute in container batches with varying system load and timing constraints.
Through Earliest Deadline First (EDF) scheduling we can reach high theoretical utilization rates of 100\%~\cite{Buttazzo2011}.
We partition resources via \textit{CGroups} so to virtually address every resource slice as if it were a separate computation unit.
The orchestration software of Section~\ref{sec:orch} will help us in this manner by managing interrupts, creating \textit{CGroups} and assigning its slices, and managing system resources to isolate them from our test containers during run-time.
First, we observe performance variation by changing kernel boot parameters of the off-the-shelf OS chosen in the latency tests.
During multiple reboots, we apply boot time kernel settings such as scheduler tick timing, scheduler isolation and RCU back-off CPUs.
The goal is to find settings that promise steady execution on the three hardware instances.
A stable median, low average and standard deviation indicate ideal kernel configurations for each machine type. 
Next, we compare the performance of on the three test architectures with the most stable setup.
We increase and mix task configurations, and verify the testing run-time determinism in long-term execution.
Dropping of performance and the amount missed deadlines ascertain then the absolute upper sharing bound.


During both experiments we work to identify parameters of Equation~\ref{eq:deadline}.
The latency tests focus on grasping the firing time of a task, $f_i$, using low footprint capturing software and logs. First we test in idle, then using stress on CPU, I/O and random memory access.
The performance tests try to identify the noise the OS, the hypervisor and concurrently running system tasks cause, $n_ij$ in Equation~\ref{eq:deadline}. 
Our test model relies on calibration loops to reproduce similar load scenarios for all three system instances, fixing the task time $t_i$.
We remove control task related noise and delays ($n_ik$) by reducing the test tool to computation only.
The test software locks its memory in a page, avoids IPC or I/O, and reduces involuntary task switches where possible.
For all data we then produce statistics containing minimum, maximum, average and standard deviation.
Further statistics such as median skew and test group maximum values are added to the performance tests to allow an assessment of distribution and uniformity.


\subsection{Context evaluation}

\begin{table}
	\centering
	\begin{threeparttable}
		\caption{Summary of tested operating systems}        
		\begin{tabular}{ p{3.5cm} p{3cm} p{3cm} p{3cm} }
			\headrow
			\thead{Name} & \thead{Pros} & \thead{Cons} & \thead{Evaluation} \\
			resinOS + patch(X3/PRT)& Small, automatic updates, Balena pre-installed & manual patch, possible efficiency issues & Maintenance high, may not work at max efficiency\\
			Ubuntu Core + patch (X3/PRT)& Small, automatic updates & manual patch, possible efficiency issues & Maintenance high, may not work at max efficiency \\
			Ubuntu LTS + Xenomai 3 & Ubuntu network and technologies, Separation of RT and nRT & Recompiling of applications needed, many RT cores drops performace & Expect high performance, medium maintenance\\
			Ubuntu LTS + PREEMT\_RT & Ubuntu network and technologies, up-to date & Keep awareness of limits, driver and hardware management needed & Med-high performance and low maintenance expected \\
			\hline
		\end{tabular}
		\label{tab:oss}
	\end{threeparttable}
\end{table}	

We first explore the running context to asses possible system software candidates for the migration.
We review state-of-the-art operating systems that can provide both (hard) real-time and container framework support. 
Besides OSs targeted for server infrastructures, we also evaluate some lightweight OS.
The selected OS must exploit the given resources properly, allowing the hardware to perform at its best while not increasing the burden of operation.
Container daemons are selected based on features, ease of use, maintenance and system footprint.

In Hofer et al.~\cite{Hoferetal2019} we evaluated Containerization techniques and OSs for our tests. 
After manual tryouts and specification evaluations, we selected \emph{Balena}\footnote{~Balena project homepage: https://www.balena.io}, or the more featured container engine \emph{Docker}\footnote{~Docker homepage: https://www.docker.com} as virtualization technique of choice.
Table~\ref{tab:oss} summarizes the operating system selection under review.
Overall, we identified Ubuntu Server LTS with the PREEMPT\_RT patch as the most promising approach.
It gives the best compromise of low maintenance and possible achievable performance.
Ubuntu Server LTS with the Co-Kernel based Xenomai 3 patch results second due to some scalability and maintenance issues.
For further details on OS and Containerization choices, refer to Sections~V and~VI of the related paper.

Based on these results, we need to use two different Linux kernel versions for latency and performance tests.
As we want to compare latency on Xenomai and PREEMPT\_RT patches systems, we choose for the first experiment the Linux kernel version 4.9.51, the latest available release featuring both patches at time of test.
The performance based tests however require newer systems that include the EDF scheduler and Greedy Reclamation of Unused Bandwidth (GRUB) algorithm, available only in kernel versions 4.13 or higher. 
These operate thus on the latest Ubuntu LTS release and patch, i.e. Ubuntu Server 18.04.2 and \texttt{PREEMPT-RT} 4.19.50-rt22. 
Both kernel versions with patches can be build and restored for all three architectures using an automated script, available online~\cite{homep01}. 

\section{Tests for latency and optimization}
\label{sec:tests}

\subsection{Experiment Setup}

For our tests we have chosen the following systems. 
The bare-metal server features two Intel Xeon X5560 (Q1'09) processors on 8 cores, 16 threads, limited to two cores for our experiments.
For hypervisor based tests, we selected Amazon Web Services (AWS) to host the cloud-based environments.
Their recent virtual instances use a new hypervisor based on KVM, called \texttt{HVM}, which allows direct assignment and control of hardware and resources reducing the virtualization overhead.
The new instances offer comparable HPC performance, but greater flexibility and scalability~\cite{amazon01}.

We selected an AWS HVM Type 1 hypervisor based T3.xlarge generic and a C5.xlarge computation optimized instance.
The T3 instances feature an Intel Xeon Platinum 8100 or 8200 (Q3'17/Q2'19) series CPU while C5 run a custom Intel Xeon Gold 6200 series (Q3'17) CPU. 
A typical T3 instance is further limited to a 40\% CPU baseline. If an instance exceeds that level of CPU, it will eventually be throttled down to 40\%. 
An ``Unlimited'' enabled variant, yet, allows for CPU bursts up-to 100\% at an additional price. 
Both AWS instances run on 4 virtual CPUs, 8 or 16 GB of RAM, shared resources and use a custom configured kernel set-up to support their proprietary hardware in Ubuntu.
%

\subsection{System latency tests}
\label{sec:virtsust}

We use \textit{cyclictest}~\cite{rttests01} Version 1.0 to measure the latency of cyclic firing behavior of a real-time application and \textit{stress}~\cite{stress02} to simulate load in the system.~\footnote{\emph{stress} has been replaced recently by stress-ng, a newer version with more features.}
The offline preliminary tests run on a dual core, 4 thread, i7 Skylake (U) system, while the main hardware comparison tests run on the three systems detailed in Section~\ref{sec:method}.
During the progressive isolation of CPU resources we measure the idle firing time and firing time change with every CPU runs stressing threads.
Once found the best setup we perform one idle and one stressed test for each configuration.
All variants, i.e., Standard Ubuntu, Xenomai and PREEMPT-RT patch run the tests for at least one million firing loops.
The logged results are then used for the long term test evaluation.

Further details, the script executing all the tests, the installation scripts, the experiment data and technical details and results are available in ~\cite{Hoferetal2019} and online~\cite{homep01}.

\subsubsection{Execution and Results}

\begin{figure}[tb]
	\centering
	\includegraphics[width=0.7\linewidth]{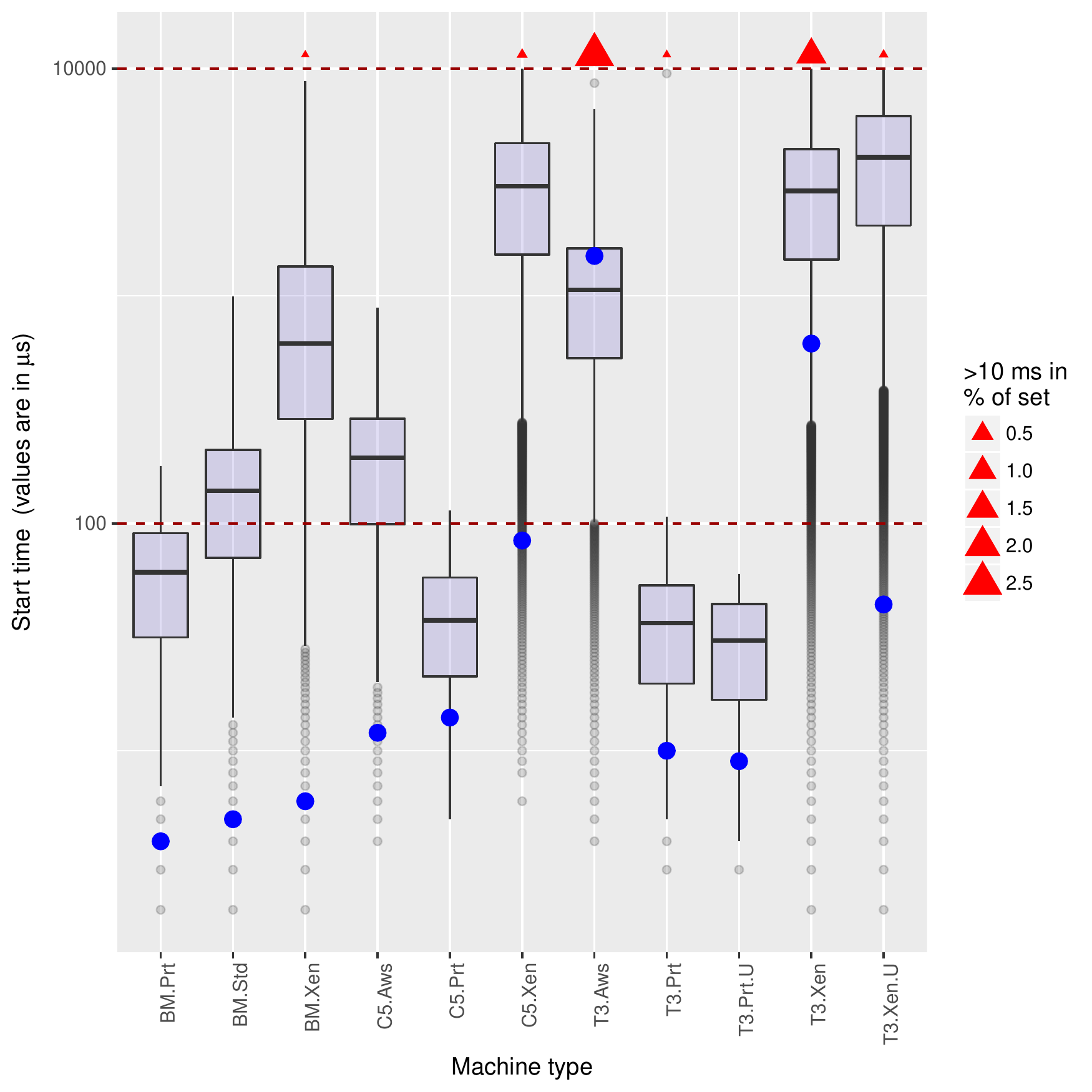}
	\caption{Boxplot of latencies, with mean (blue dot) and overshoot size. 
		The represented machine types on the x-axis are bare-metal (BM), AWS T3 (T3) and AWS C5 (C5) with corresponding OS patch (Std, Standard;Prt, PREEMPT\_RT patched; AWS, AWS Standard; and Xen, Xenomai patched). 
		Machine types with an~\emph{.U} in their tag refer to Unlimited enabled systems, allowing small CPU bursts for a surcharge. 
	}
	\label{fig:plot}
\end{figure}

In summary, the latency tests give the following main results.
The first preliminary latency tests determined that for our purposes, guest-host CPU isolation with load balancer is the best setting.
Table~I in~\cite{Hoferetal2019} displays test results for the preliminary test.
Figure~\ref{fig:plot} then shows the comparison test results with our found best setting. 
Ideally, the maximum firing delay of the threads should stay below $\frac{1}{10}^{th}$ of the cycle time, which we assumed to be $100ms$ for sake of comparison in this study.
Therefore, Figure~\ref{fig:plot} features two reference lines visualizing the boundaries for typical thresholds, one at $10ms$ (for a $100ms$ cycle) and $100\mu s$ (for a $1ms$ cycle).

A total of ten million loops over multiple hours have been executed for each configuration.
All results obtained have been gathered under \textit{stress} and should be considered the worst case scenarios.
Among all standard kernel configurations, the reference bare-metal solution equipped with any of the three patches (BM, left box-plots in Figure \ref{fig:plot}) performs best in mean. 
If we consider the PREEMPT\_RT configurations (Prt) across all machine types, the bare-metal set-up performs best in mean but not in spread as the box-plot whisker spans higher and almost reaching the  $100\mu s$ threshold.
With only 96 occurrences out of 10 million (0.00096\%) exceeding the upper limit, a general T3  instance with PREEMPT\_RT can be an economic solution for a bare-metal replacement where strict determinism is not needed or cycle times are higher than the peak value measured, $49ms$.
It shows the lowest spread and peak ($114\mu s$) among the measured instances that only a PREEMPT\_RT T3-Unlimited enabled unit outperforms.

\subsubsection{Latency results discussion}

\textit{\textbf{RQ1:} What are possible off-the-shelf system configurations that make resource sharing through containers viable?}

\noindent
Among the examined low maintenance options we identified Ubuntu 16.04 LTS with the PREEMPT\_RT real-time patch and Docker containers as best fit.
By observing systems under stress and analyzing task latency across different configurations, we came up with four different solutions suitable for migration to application containerization.
These solutions maintain wake-up determinism at different levels as follows:
\vspace{-10px}
\begin{enumerate}
	
	\item The bare-metal solution (BM) ensures the most deterministic behavior for hard real-time requirements. 
	Even though it is the weakest among all configurations in terms of CPU resources, the strict bond between hardware and software boosts its responsiveness.
	
	\item The virtualized instance C5 with PREEMPT\_RT patch is the best non-hardware solution for hard real-time requirements that trades off good average latency and deterministic behavior. 
	While it still suffers from some Hypervisor latency, the exclusiveness of CPU access and the ability to control C-states allow reducing non I/O induced noise and plot better value consistency.
	
	\item The T3 unlimited instance with PREEMPT\_RT is a cheap solution with good average latency.
	As there is no guarantee on the availability or responsiveness of extra CPU power, these configurations can be chosen as an intermediate solution between T3 and C5 instances. 
	
	\item The T3 instance with PREEMPT\_RT is a viable solution with good average latency that might not qualify for hard real-time requirements.
	Also, this T3 instance may not ensure the physical CPU exclusiveness. For this reason, the C5 PREEMPT\_RT instance may be a better choice for stricter timing requirements. 
	
\end{enumerate}

\vspace{-10px}
\noindent In conclusion, the results are promising and confirm the feasibility of migration to IAAS solutions.

\subsection{Performance tests for resource optimization with container orchestration}
\label{sec:orchestr}

We perform the following resource efficiency to tests by placing a set of real-time applications on shared resources.
For this purpose, we use the real-time test software \texttt{rt-app}~\cite{rtapp01} to create configurable dummy applications.
We place them into separate containers and configure them with running periods and computation times.
For simplicity, we match relative computation deadline $d_i$ and period $p_i$ of a container $i$ in all tests (See Equation~\ref{eq:deadline}).
Beyond the worst case value expected for $r_i$, WCET $w_i$, the app configuration requires a simulated run-time parameter $t_i$.
The latter defines the time the application spends performing dummy loop calculations, as an approximation to the supposed run-time of the simulated task.
As the amount loops to perform depends on a constant value set at a startup (calibration) and we isolated task caused noise, the resulting execution time $c_i$ depends solely on interaction with the system.
During test execution, the orchestrator and \emph{rt-app} monitor run-time behavior of the system.
With these results we can detect of startup latency, execution jitter and deadline misses.

Each test batch consists of the following four configurations:
\begin{enumerate}[label=Test Case~\arabic*,start=1]
    \item ~--~\textit{lower bound:} homogeneous period and run-time among all containers executing on the same resources with a WCET $w_i$ smaller than the best case scheduler's wake-up granularity ($1000\mu s$). 
    With this test, we force high resolution granularity scheduling, causing more scheduler calls than planned for the highest scheduler tick rate.
   	The test setup consists of ten containers with a WCET $w_i$ of $900\mu s$.
   	With a period and deadline of $10ms$ each, this results in a resource utilization factor $U$ of $0.9$.
    \item ~--~\textit{upper bound:} homogeneous period and run-time among containers executing on the same resources, with a run-time to period ratio $(\frac{t_i}{p_i})$ close or equal to $0.5$.
    The kernel limits the schedulers refresh rate to $1000/4000\mu s$ for 1000/250Hz systems, making this configuration a scheduling challenge.
    The configured test case includes two containers with $2.5ms$ WCET and $5ms$ period.
    \item ~--~\textit{diversity:} mixed periods and run-times for each container executing on the same resources.
    Irregular run-times should challenge the possibility of execution alignment where containers always run in the same order. 
    Moreover, the deadline priority continues to rotate, helping to determine stability in mixed scenarios.
    This test case consists of a mixed set of containers: one container set to $2.5/5ms$, one $900\mu s/10ms$, and one configured as $3/9ms$ for worst case computation time and deadline/period, respectively.
	\item ~--~\textit{Simulation test:} As a conclusive test, we emulate the scenario of our example application. 
	Trough parallel operation of multiple instances of the flow control software, we verify the boundaries for this use case.
	The test configuration includes ten containers with the period and run-time homogeneous among all tasks and running on the same resources. 
	The timing is based on the values of the example in Section~\ref{sec:bgnd} ($10ms$ run-time, $100ms$ deadline and period).
\end{enumerate}

The test scripts and complete test results can be found in the project repository~\cite{homep02}.

\subsubsection{Execution and results}

The first test batch for kernel configuration shows the full dynamic tick configuration with mandatory back-off as performing best.
The setup scored the best results in most runs for average and median stability.
Test case 1's values gave standard deviations of $31\mu s$ on bare-metal, $7\mu s$ on type C5 and $15\mu s$ on type T3 instances when running ten containers in parallel. 
Second by performance and stability is the fixed-tick kernel configuration. 
This second configuration turns out useful if more than one task is available to run or next in line at the same time, mostly when mixed with non-deadline oriented schedules.
For the long-run tests of test batch two, we thus choose a PREEMPT\_RT kernel with full dynamic ticks and RCU back-off, with a run-time of 15 minutes each.

In Test Batch Two, we repeated the same tests on all three systems.
To test the repeatability, we re-calibrated and repeated the tests multiple times.
Additional results and diagrams can be found in the project archive~\cite{homep01}.
Tables~\ref{tab:resgrp1} to~\ref{tab:resgrp4} report the results of the four test cases.
We display numbers for test case 1 and 4 with loads from close to 50 up-to 100\% of CPU time only.

\begin{table}
	\centering
	\begin{threeparttable}
		\caption{Test batch 2, test case 1}
		\begin{tabular}{l *9{c}}
			\headrow
			\thead{Configuration} & \multicolumn{3}{c}{\bfseries Bare-Metal}& \multicolumn{3}{c}{\bfseries AWS T3}& \multicolumn{3}{c}{\bfseries AWS C5}\\
			& \thead{AVG} & \thead{SKW} & \thead{SD\_MX}& \thead{AVG} & \thead{SKW} & \thead{SD\_MX}& \thead{AVG} & \thead{SKW} & \thead{SD\_MX}\\
			4 units <50\% & 935 & 0/14 & 22.83 & 903 & 3/11 & 16.21 & 914 & 2/2 & 5.05\\
			5 units <60\%  & 950 & 0/12 & 23.91 & 904 & 4/11 & 16.33 & 913 & 0/0 & 6.53\\
			6 units <70\%  & 969 & 0/6 & 11.34 & 905 & 5/11 & 15.73 & 913 & 0/4 & 5.97\\
			7 units <80\% & 930 & 0/19 & 22.37 & 904 & 1/11 & 16.46 & 913 & 0/4 & 5.48\\
			8 units <90\% & 926 & 0/6 & 12.28 & 904 & 2/10 & 15.43 & 913 & 1/3 & 5.43\\
			9 units <100\% & 920 & 0/13 & 23.33 & 904 & 4/10 & 15.14 & 914 & 1/3 & 5.12\\
			10 units $\approx$100\% & 933 & 0/6 & 11.66 & 904 & 4/9 &  14.81 & 914 & 1/4 & 5.50\\
			\hline  
		\end{tabular}
		\label{tab:resgrp1}
	\end{threeparttable}

	\bigskip
	
	%
	\begin{threeparttable}
		\caption{Test batch 2, test case 1}
		\begin{tabular}{l *9{c}}
			\headrow
			\thead{Configuration} & \multicolumn{3}{c}{\bfseries Bare-Metal}& \multicolumn{3}{c}{\bfseries AWS T3}& \multicolumn{3}{c}{\bfseries AWS C5}\\
			& \thead{AVG} & \thead{SKW} & \thead{SD\_MX}& \thead{AVG} & \thead{SKW} & \thead{SD\_MX}& \thead{AVG} & \thead{SKW} & \thead{SD\_MX}\\
			1 unit <60\%  & 2584 & 4 & 21.28 & 2510 & 12 & 27.34 & 2538 & 1 & 6.48\\
			2 units $\approx$100\% & 2521 & 13* & 23.23 & 2506 & 14* & 26.03 & 2535 & 0* & 5.75\\
			\hline  
		\end{tabular}
		\label{tab:resgrp2}
	\end{threeparttable}

	\bigskip

	\begin{threeparttable}
		\caption{Test batch 2, test case 3.}
		\begin{tabular}{l *9{c}}
			\headrow
			\thead{Configuration} & \multicolumn{3}{c}{\bfseries bare-metal}& \multicolumn{3}{c}{\bfseries AWS T3}& \multicolumn{3}{c}{\bfseries AWS C5}\\
			& \thead{AVG} & \thead{SKW} & \thead{SD\_MX}& \thead{AVG} & \thead{SKW} & \thead{SD\_MX}& \thead{AVG} & \thead{SKW} & \thead{SD\_MX}\\
			1 unit <60\% & 2579 & 10 & 22.16 & 2507 & 9 & 23.23 & 2534 & 1 & 7.70\\
			2 units <90\%  & 2589 & 4/14 & 22.65 & 2569 & 4/7 & 19.01 & 2587 & 0/1 & 5.40\\
			3 units $\approx$100\% & 2269 & 1/7 & 26.54 & 2179 & 7/37 &  37.25 & 2183 & 0/1 & 11.13\\
			\hline  
		\end{tabular}
		\label{tab:resgrp3}
	\end{threeparttable}

	\bigskip

	%
	\begin{threeparttable}
		\caption{Test batch 2, test case 4.}
		\begin{tabular}{l *9{c}}
			\headrow
			\thead{Configuration} & \multicolumn{3}{c}{\bfseries Bare-Metal}& \multicolumn{3}{c}{\bfseries AWS T3}& \multicolumn{3}{c}{\bfseries AWS C5}\\
			& \thead{AVG} & \thead{SKW} & \thead{SD\_MX}& \thead{AVG} & \thead{SKW} & \thead{SD\_MX}& \thead{AVG} & \thead{SKW} & \thead{SD\_MX}\\
			4 units <50\% & 10712 & 0/8 & 31.78 & 10072 & 14/16 & 45.69 & 10139 & 2/2 & 12.03\\
			5 units <60\%  & 10614 & 8/10 & 35.66 & 10056 & 14/16 & 35.46 & 10310 & 1/3 & 68.98\\
			6 units <70\%  & 10132 & 3/10 & 34.87 & 10038 & 4/7 & 23.30 & 10136 & 1/2 & 10.55\\
			7 units <80\% & 10115 & 0/11 & 31.25 & 10166 & 49/57 & 123.36 & 10138 & 1/2 & 11.08\\
			8 units <90\% & 10104 & 3/12 & 32.20 & 10052 & 10/16 & 41.55 & 10137 & 1/2 & 9.04\\
			9 units <100\% & 10356 & 3/8 & 29.83 & 10059 & 5/16 & 126.17 & 10138 & 1/2 & 9.94\\
			10 units $\approx$100\% & 10089 & 4/10 & 46.29 & 10027 & 1/3 &  12.75 & 10136 & 1/2 & 9.91\\
			\hline  
		\end{tabular}
		\label{tab:resgrp4}
	\end{threeparttable}

	\begin{tablenotes} 
		\item Performance variations with an increasing number of containers. Values in $\mu s.$
		\item AVG - average run-time of container set over all measured run-times
		\item SKW - absolute min and max distance between average run-time value and median for the configuration
		\item SD\_MX - standard deviation of the container with the highest skew 
		\item \textbf{*} The starred values indicate runs where at least one thread did not produce log output
	\end{tablenotes}

\end{table}			

\begin{enumerate}[label=Test Case~\arabic*:,start=1]
	\item
	Even though in these tests the system load never reaches 100\%, the table shows continuity among all instances with minor variations of about one to five percent.
	The AWS C5 instance performed best among the three candidate systems.
	With a steady average run-time, close to null skew and an unvarying standard deviation, the most resourceful of the three keeps a steady and deterministic run.
	The other virtual instance is slightly slower but nonetheless keeps a small variation bound.
	Interestingly, in this configuration the bare-metal system shows the most jitter.
	Different from T3 and C5, the values of skew and standard deviation halve and double from test to test, having a similar varying impact on the average run-time.
	In the highest configuration, systems reach loads of 90 to 93\%. 
	No runs across all configurations show any overshoot, confirming the feasibility of handling multiple hard real-time containers while suffering from a relatively small system noise.

	\item 
	This test case contains only two configurations: one or two containers.
	Unfortunately, the system noise is already high enough to make a single container exceed 50\% of CPU load. 
	A consequent phenomenon is that configuration two produces only one run time log on all candidate systems.
	All performed tests do not deliver enough values to get minimum and maximum for the skew and deviation, marked by an asterisk in Table~\ref{tab:resgrp2}, leaving a doubt on possible performance change.
	However, the visible data does suggest a small skew of the distribution, but the standard deviation tends to remain low.
	The high amount deadline misses for configuration two, around 20 or more for all systems, nonetheless upholds that system overload causes the expected misbehavior.

	\item 
	The average run-time in this mixed container spans all running containers and serves as variation monitor rather than a proper average.
	Indeed, the high numbers of deadline misses in this third configuration causes task preemption and run-time values to boost.
	Like in the previous test cases, AWS C5 keeps a steady and centered distribution displaying no skew. 
	The bare-metal configuration behaves similarly to test case 1 with a slightly trembling skew around a tenth of microsecond while maintaining a rather constant standard deviation. 
	The general purpose instance however suffers from the high load of the last test and drifts into an unusually high skew and standard deviation of $37\mu s$.
	Test 1 and 2 already use close to 90\% of CPU time, causing first deadline overshoots for bare-metal (20157) and AWS T3 (25) in configuration two.
	The AWS C5 instance instead does not fail any run in this first two tests. 
	For test three, yet, all configurations show misses in the order of thousands in this 15 minute test.
	
	\item 
	Also in this fourth test case, the AWS C5 keeps the distribution centered to the median and, except for the second test, stays with distributions close to $10\mu s$.
	Similarly, the bare-metal system shows the same small distribution skew and slight jitter in its standard deviation as it did for case 1.
	The AWS T3 however, in this configuration seems to suffer more from system latency and noise, showing higher skew and standard deviations than in any test before.
	As for maintaining deadlines, the results keep stable for all tests except full load where they show 3, 244 and 3 overshoots for bare-metal, T3 and C5 respectively.
\end{enumerate}

\subsubsection{Performance results discussion}

\textit{\textbf{RQ2:} What is the achievable level of CPU sharing with a standard real-time enabled kernel?}

\noindent
The static orchestration tests highlight some additional facts beyond the best performing configurations.
One thing that arose is the performance increase of generic instances (T3/T3U) when moving from 10 to 50\% of CPU load. 
This improvement is like those observed for latency tests, probably due to the hypervisors CPU management. 
It may move the vCPU thread or fill the remainder with other instance's vCPU threads.
Different from computation intensive instances (C5), a type T3 does not require locking onto a specific physical CPU, adding volatility and latency.
The latency witnessed during low system load may thus be a product of virtual thread shifting.
As a T3 instance has no hardware access, this CPU could continuously transition to a lower C state, adding more latency for the thread to be back in execution.
Thus, similarly to the conclusions of the latency tests in Section~\ref{sec:virtsust}, a higher CPU load reduces system induced latency and noise.

Another observation is that with high CPU load, the real-time tasks take over minor threads. 
This causes misbehavior such as incomplete or empty log files, tasks not terminating at predefined times or sometimes unresponsiveness of the system. 
Furthermore, in some cases, one container in the test configuration kept a continuous run-time deviation from the preset run-time. 
Occasionally, the average and median run-time kept around $12ms$ instead of $10ms$, while displaying same jitter and deviation behavior.
Despite this deviation, the stability and thus determinism of the run-time values are of no hindrance to hard real-time operation. 

During preliminary testing, we noted that the restart of a virtual instance on the AWS cluster causes it to move to a different system rack. 
Given that hardware across system racks may not be equal, e.g., Xeon 8100 vs 8200 series CPU, this change after shutdown is a variable to be considered.
While this influenced the calibration for AWS based tests, it does not influence the comparison among the results of the same virtual instance.

The resulting run-time data from both test batches shows that resource sharing for real-time containers is feasible.
Properly configured, a system can reach a utilization limit of $0.9$ or $90\%$.
Through our tests we have shown that, although under stress, both latency and determinism reach desired values.
%
Among all, the AWS C5 shows the most stable run-time values. 
It is the most resourceful of all systems and thus likely suffering the least from background noise. 
Being virtual, it does not respond directly to hardware interrupts like the bare-metal system, softening amount and duration of interrupts.
However, this does not mean it is not influenced by system noise.
As seen in test configuration two of Table~\ref{tab:resgrp4}, the C5 system can still be subject to major variations. 
The bare-metal instance, yet, shows higher fluctuation in skew and standard deviation, but still stays steady in a certain range.
In all tests, the results for skew and deviation remained within $20$ or $30\mu s$.
While this jitter may seem a problem, the fact that it can be isolated to this range make it predictable and thus ideal for hard real-time use.
Lastly, the generic AWS T3 shows the worst but still rather stable run-time behavior.
The highest fluctuations are shown in test case 4, where idle times between cycle repetition are the longest. 
Indeed, this confirms that during idling, the hypervisor may change the physical CPU reservation. 
If we consider these constraints, also an economic generic AWS T3 instance may suffice our computational needs. 

In the end, all systems show adequate stability for the sample loads we created.
The worst variation of system run-time stays within $126 \mu s$, a value that has to be considered when dealing with hard deadlines in the order of few milliseconds or less. 
However, this confirms that all setups allow shared computational loads up-to and exceeding 90\%.  
Only close to full load the systems starts to suffer from a deadline overshoots.
Starting from these results, we can now investigate if off-the-shelf technology keeps the process viable once task I/O ($n_ik$) and network latency are taken into account.

\section{Lessons learned}

Thanks to the performed test series and after discussing results and consequences, we have drawn few lessons we learned that could be beneficial for practitioners aiming to use application containers for industrial control as in the following:

\begin{itemize}    

    \item[Real-time requirements, and consequently architecture, are application specific.]
	While a generic architecture as presented in Section~\ref{sec:arch} covers most situations, the current layout changes for each case.
	Like in the motivating example, Figure~\ref{fig:smpcase}, levels may merge where the environment requires it, and give the final architecture a different, reduced shape. However, responsibilities, function and goals remain unchanged.
    
	\item[Picking a real-time capable OS does not guarantee determinism.]
	Different OSs have distinct trade-offs. 
	While the Linux Xenomai patch outperforms the PREEMT\_RT patch, its induced kernel overhead limits systems scalability.
	When choosing OS, we have to closely match hardware and constraints for best results.

	\item[Modern virtualization techniques perform well enough to accommodate hard real-time environments.]
	Both, the latency and performance test showed satisfying results confirming viability of an application migration.
	Depending on task configuration, we can reach subscription rates exceeding 90\% of CPU resources.
	The next and last constraint to tackle will be the network and I/O latency.
	This, however, depends on the applications' timing requirements and thus, needs further investigation. 

	\item[Direct hardware access decreases latency and improves responsiveness.]
	Despite the less powerful hardware, the Bare-Metal server still outperforms newer Hypervisor based instances for task responsiveness.
	Similarly, limited access to CPU resources improves virtualized performances, i.e. AWS C5 vs T3-Unlimited.
	Thus, although possible, virtualized instances require newer and better hardware to reach similar performance.
	A practitioner might thus need to consider resource sharing beyond control containers to reduce hardware installation costs.
	The architecture of Section~\ref{sec:arch} helps to address this job. 

	\item[Economic virtual instances may suffice for less strict determinism requirements.]
	Generic AWS T3 shared instances show comparable results for task firing latency, but add variability when under stress.
	While this variability discourages their use in environments with strict timing requirements, i.e., task periods of few milliseconds or less, it enables them however for less critical operations, e.g., periods in 100's of milliseconds like in the motivating example, Section~\ref{sec:bgnd}.

\end{itemize}

\section{Conclusions and Future Work}
\label{sec:conclusion}

In this paper, we explored limits and feasibility of migrating real-time applications from bare-metal servers to virtualized IAAS configurations.
We showed that containerization offers a novel paradigm for control applications. 
Previously isolated computation tasks, however, may operate concurrently and interact with each other, potentially influencing timing performance.
We concur with Goldschmidt \textit{et al.}~\cite{Goldschmidtetal2018} that this new paradigm requires investigation on topics such as container security, restricted access and intra-container data exchange. 
We suggest an architecture to help migration and placement of these new applications in an Industry 4.0 focus.
Through the alignment of technologies and the interconnection of attributes, the proposal enables features previously not available.
We next introduced an orchestration tool that can schedule real-time containers based on pre-configured capacities.
We showed configurations that maximize resource utilization without significantly impacting overall execution determinism.
Through targeted tests, we verified migration viability and influence on a computation only task considering system I/O and latency. 

In future work, I/O and system latency will be investigated and dynamic allocation strategies will be exploited to further improve system performance.
A dynamic orchestration algorithm will help to tackle issues that arise when task do not respect their designed parameters.
This new configuration will also help to increase robustness of a system and detect a deviation of task behavior due to cyber-attacks or externally induced overloads.
New latency and performance tests on industrial use cases will help further analyze limits and possibilities for shared-resource real-time systems, including robustness and behavior when under attack.



\bibliography{bibliography}

\begin{thebibliography}{36}
\providecommand{\natexlab}[1]{#1}
\providecommand{\url}[1]{\texttt{#1}}
\providecommand{\urlprefix}{}

\bibitem[{Telschig et~al.(2018)Kilian Telschig and Andreas Sch\"onberger and
  Alexander Knapp}]{Telschigetal2018}
Telschig K, Sch\"onberger A, Knapp A.
\newblock A Real-Time Container Architecture for Dependable Distributed
  Embedded Applications.
\newblock In: 2018 {IEEE} 14th International Conference on Automation Science
  and Engineering ({CASE}) {IEEE}; 2018. .

\bibitem[{Hofer(2018)Florian Hofer}]{Hofer2018}
Hofer F.
\newblock Architecture, technologies and challenges for cyber-physical systems
  in Industry 4.0 - A systematic mapping study.
\newblock In: 12th ACM / IEEE International Symposium on Empirical Software
  Engineering and Measurement (ESEM); 2018. .

\bibitem[{Tasci et~al.(2018)Timur Tasci and Jan Melcher and Alexander
  Verl}]{Tascietal2018}
Tasci T, Melcher J, Verl A.
\newblock A Container-based Architecture for Real-Time Control Applications.
\newblock In: 2018 {IEEE} International Conference on Engineering, Technology
  and Innovation ({ICE}/{ITMC}) {IEEE}; 2018. p. 1--9.

\bibitem[{Moga et~al.(2016)Moga, Alexandru and Sivanthi, Thanikesavan and
  Franke, Carsten}]{Mogaetal2016}
Moga A, Sivanthi T, Franke C.
\newblock {OS}-level virtualization for industrial automation systems: Are We
  There Yet?
\newblock In: Proceedings of the 31st Annual {ACM} Symposium on Applied
  Computing - {SAC} '16 {ACM} Press; 2016. p. 1838--1843.

\bibitem[{Goldschmidt and Hauck-Stattelmann(2016)Thomas Goldschmidt and Stefan
  Hauck-Stattelmann}]{GoldschmidtHauck-Stattelmann2016}
Goldschmidt T, Hauck-Stattelmann S.
\newblock Software Containers for Industrial Control.
\newblock In: 2016 42th Euromicro Conference on Software Engineering and
  Advanced Applications ({SEAA}) {IEEE}; 2016. p. 258--265.

\bibitem[{Fazio et~al.(2016)Maria Fazio and Antonio Celesti and Rajiv Ranjan
  and Chang Liu and Lydia Chen and Massimo Villari}]{Fazioetal2016}
Fazio M, Celesti A, Ranjan R, Liu C, Chen L, Villari M.
\newblock Open Issues in Scheduling Microservices in the Cloud.
\newblock {IEEE} Cloud Computing 2016 sep;3(5):81--88.

\bibitem[{Lee et~al.(2015)Jay Lee and Behrad Bagheri and Hung-An
  Kao}]{Leeetal2015}
Lee J, Bagheri B, Kao HA.
\newblock A Cyber-Physical Systems architecture for {Industry} 4.0-based
  manufacturing systems.
\newblock Manufacturing Letters 2015 jan;3:18 -- 23.

\bibitem[{Telschig and Knapp(2017)Kilian Telschig and Alexander
  Knapp}]{TelschigKnapp2017}
Telschig K, Knapp A.
\newblock Towards Safe Dynamic Updates of Distributed Embedded Applications in
  Factory Automation.
\newblock In: 22nd {IEEE} International Conference on Emerging Technologies and
  Factory Automation ({ETFA}) No. reconfiguration., {IEEE}; 2017. p. 1--4.

\bibitem[{Wu et~al.(2017)Dazhong Wu and Shaopeng Liu and Li Zhang and Janis
  Terpenny and Robert X. Gao and Thomas Kurfess and Judith A.
  Guzzo}]{Wuetal2017}
Wu D, Liu S, Zhang L, Terpenny J, Gao RX, Kurfess T, et~al.
\newblock A fog computing-based framework for process monitoring and prognosis
  in cyber-manufacturing.
\newblock Journal of Manufacturing Systems 2017 apr;43:25--34.

\bibitem[{Terrissa et~al.(2016)Labib Sadek Terrissa and Safa Meraghni and Zahra
  Bouzidi and Noureddine Zerhouni}]{Terrissaetal2016}
Terrissa LS, Meraghni S, Bouzidi Z, Zerhouni N.
\newblock A new approach of {PHM} as a service in cloud computing.
\newblock In: NA, editor. 2016 4th {IEEE} International Colloquium on
  Information Science and Technology ({CiSt}) IEEE, {IEEE}; 2016. p. 610--614.

\bibitem[{Schroeder et~al.(2016)Greyce Schroeder and Charles Steinmetz and
  Carlos Eduardo Pereira and Ivan Muller and Natanael Garcia and Danubia
  Espindola and Ricardo Rodrigues}]{Schroederetal2016}
Schroeder G, Steinmetz C, Pereira CE, Muller I, Garcia N, Espindola D, et~al.
\newblock Visualising the digital twin using web services and augmented
  reality.
\newblock In: NA, editor. 2016 {IEEE} 14th International Conference on
  Industrial Informatics ({INDIN}) {IEEE}; 2016. p. 522--527.

\bibitem[{Roy et~al.(2016)R. Roy and R. Stark and K. Tracht and S. Takata and
  M. Mori}]{Royetal2016}
Roy R, Stark R, Tracht K, Takata S, Mori M.
\newblock Continuous maintenance and the future {\textendash} Foundations and
  technological challenges.
\newblock {CIRP} Annals 2016;65(2):667--688.

\bibitem[{Goldschmidt et~al.(2018)Goldschmidt, Thomas and Hauck-Stattelmann,
  Stefan and Malakuti, Somayeh and Gr{\"u}ner, Sten}]{Goldschmidtetal2018}
Goldschmidt T, Hauck-Stattelmann S, Malakuti S, Gr{\"u}ner S.
\newblock Container-based architecture for flexible industrial control
  applications.
\newblock Journal of Systems Architecture 2018;84:28--36.

\bibitem[{Garc{\'{\i}}a-Valls et~al.(2014)Marisol Garc{\'{\i}}a-Valls and
  Tommaso Cucinotta and Chenyang Lu}]{Garcia-Vallsetal2014}
Garc{\'{\i}}a-Valls M, Cucinotta T, Lu C.
\newblock Challenges in real-time virtualization and predictable cloud
  computing.
\newblock Journal of Systems Architecture 2014 oct;60(9):726--740.

\bibitem[{Hallmans et~al.(2015)Hallmans, Daniel and Sandström, Kristian and
  Nolte, Thomas and Larsson, Stig}]{Hallmansetal2015}
Hallmans D, Sandström K, Nolte T, Larsson S.
\newblock Challenges and Opportunities when Introducing Cloud Computing into
  Embedded Systems; 2015. p. 454--459.

\bibitem[{Felter et~al.(2015)Wes Felter and Alexandre Ferreira and Ram Rajamony
  and Juan Rubio}]{Felteretal2015}
Felter W, Ferreira A, Rajamony R, Rubio J.
\newblock An updated performance comparison of virtual machines and Linux
  containers.
\newblock In: 2015 {IEEE} International Symposium on Performance Analysis of
  Systems and Software ({ISPASS}) {IEEE}; 2015. p. 171--172.

\bibitem[{Arango et~al.(2017)Carlos Arango and R\'emy Dernat and John
  Sanabria}]{Arangoetal2017}
Arango C, Dernat R, Sanabria J.
\newblock Performance Evaluation of Container-based Virtualization for High
  Performance Computing Environments; 2017, arXiv preprint arXiv:1709.10140.

\bibitem[{Abeni et~al.(2018)Abeni, Luca and Balsini, Alessio and Cucinotta,
  Tommaso}]{Abenietal2018}
Abeni L, Balsini A, Cucinotta T.
\newblock Container-Based Real-Time Scheduling in the Linux Kernel.
\newblock In: Embedded Operating System Workshop 2018 (EWiLi’18); 2018. .

\bibitem[{Buttazzo(2011)Buttazzo, Giorgio C}]{Buttazzo2011}
Buttazzo GC.
\newblock Hard real-time computing systems: predictable scheduling algorithms
  and applications, vol.~24.
\newblock Springer Science \& Business Media; 2011.

\bibitem[{Hofer and Russo(????)Hofer, Florian and Russo,
  Barbara}]{HoferRusso2020}
Hofer F, Russo B.
\newblock Architecture and its vulnerabilities in Smart-Lighting systems;,
  \urlprefix\url{https://www.florianhofer.it/papers/?id=6c0b6c41c039423788a965e7cb9547155b2a33bd}.

\bibitem[{Han et~al.(2014)Song Han and Miao Xie and Hsiao-Hwa Chen and Yun
  Ling}]{Hanetal2014}
Han S, Xie M, Chen HH, Ling Y.
\newblock Intrusion Detection in Cyber-Physical Systems: Techniques and
  Challenges.
\newblock {IEEE} Systems Journal 2014 dec;8(4):1052--1062.

\bibitem[{Hofer(2018)Florian Hofer}]{Hofer2018-2}
Hofer F.
\newblock Enhancing {S}ecurity and {R}eliability for {S}mart-* {S}ystems’
  {A}rchitectures.
\newblock In: 2018 IEEE International Symposium on Software Reliability
  Engineering Workshops; 2018. p. 150--153.

\bibitem[{Gorecky et~al.(2014)Gorecky, Dominic and Schmitt, Mathias and
  Loskyll, Matthias and Zuhlke, Detlef}]{Goreckyetal2014}
Gorecky D, Schmitt M, Loskyll M, Zuhlke D.
\newblock Human-machine-interaction in the industry 4.0 era.
\newblock In: NA, editor. 2014 12th {IEEE} International Conference on
  Industrial Informatics ({INDIN}) {IEEE}; 2014. p. 289--294.

\bibitem[{iee(2019)}]{ieee01}
Time-Sensitive Networking (TSN) Task Group; 2019.
\newblock \urlprefix\url{https://1.ieee802.org/tsn/}.

\bibitem[{Robert et~al.(2012)J{\'{e}}r{\'{e}}my Robert and Jean-Philippe
  Georges and Eric Rondeau and Thierry Divoux}]{Robertetal2012}
Robert J, Georges JP, Rondeau E, Divoux T.
\newblock Minimum Cycle Time Analysis of Ethernet-Based Real-Time Protocols.
\newblock International Journal of Computers Communications {\&} Control 2012
  sep;7(4):744.

\bibitem[{Morabito(2017)Morabito, Roberto}]{Morabito2017}
Morabito R.
\newblock Virtualization on internet of things edge devices with container
  technologies: a performance evaluation.
\newblock IEEE Access 2017;5:8835--8850.

\bibitem[{Hofer et~al.(2020)Hofer, Florian and Sehr, Martin and
  Sangiovanni-Vincentelli, Alberto and Russo, Barbara}]{Hoferetal2020}
Hofer F, Sehr M, Sangiovanni-Vincentelli A, Russo B.
\newblock {ODRE} Workshop: Probabilistic Dynamic Hard Real-Time Scheduling in
  {HPC}.
\newblock In: 2020 IEEE 23rd International Symposium on Real-Time Distributed
  Computing (ISORC); 2020. .

\bibitem[{Shekhar et~al.(2018)Shekhar, Shashank and Abdel-Aziz, Hamzah and
  Bhattacharjee, Anirban and Gokhale, Aniruddha and Koutsoukos,
  Xenofon}]{Shekharetal2018}
Shekhar S, Abdel-Aziz H, Bhattacharjee A, Gokhale A, Koutsoukos X.
\newblock Performance Interference-Aware Vertical Elasticity for Cloud-Hosted
  Latency-Sensitive Applications.
\newblock In: 2018 {IEEE} 11th International Conference on Cloud Computing
  ({CLOUD}) {IEEE}; 2018. p. 82--89.

\bibitem[{Chen et~al.(2018)Chen, Jie and Cao, Chun and Zhang, Ying and Ma,
  Xiaoxing and Zhou, Haiwei and Yang, Chengwei}]{Chenetal2018}
Chen J, Cao C, Zhang Y, Ma X, Zhou H, Yang C.
\newblock Improving Cluster Resource Efficiency with Oversubscription.
\newblock In: 2018 {IEEE} 42nd Annual Computer Software and Applications
  Conference ({COMPSAC}) {IEEE}; 2018. p. 144--153.

\bibitem[{Hofer et~al.(2019)Hofer, Florian and Sehr, Martin and Iannopollo,
  Antonio and Ugalde, Ines and Sangiovanni-Vincentelli, Alberto and Russo,
  Barbara}]{Hoferetal2019}
Hofer F, Sehr M, Iannopollo A, Ugalde I, Sangiovanni-Vincentelli A, Russo B.
\newblock Industrial Control via Application Containers: Migrating from
  Bare-Metal to {IAAS}.
\newblock In: 2019 {IEEE} 11th International Conference on Cloud Computing
  Technology and Science ({CloudCom}) {IEEE}; 2019. p.~NA.

\bibitem[{Hofer(2019)Hofer, F.}]{homep01}
Hofer F, Test archive; 2019.
\newblock \urlprefix\url{http://bit.ly/2XdoYPn}.

\bibitem[{ama(2018)}]{amazon01}
Amazon AWS User guide - The C5 instance; 2018.
\newblock \urlprefix\url{https://aws.amazon.com/ec2/instance-types/c5/}.

\bibitem[{rtt(2018)}]{rttests01}
{rt}-tests - test programs for real-time kernels; 2018.
\newblock \urlprefix\url{https://directory.fsf.org/wiki/Rt-tests}.

\bibitem[{str(2018)}]{stress02}
Stress-ng - the next generation stress testing tool; 2018.
\newblock \urlprefix\url{http://kernel.ubuntu.com/~cking/stress-ng/}.

\bibitem[{rta(2019)}]{rtapp01}
rt-app - scheduler tools GitHub home; 2019.
\newblock \urlprefix\url{https://github.com/scheduler-tools/rt-app}.

\bibitem[{Hofer(2019)Hofer, Florian}]{homep02}
Hofer F, {Test archive Orchestration}; 2019.
\newblock \urlprefix\url{http://bit.ly/2PQqnJN}.

\end{thebibliography}

\end{document}